\documentclass[fleqn,usenatbib]{mnras}

\usepackage{newtxtext,newtxmath}
\usepackage[T1]{fontenc}
\usepackage{color}

\DeclareRobustCommand{\VAN}[3]{#2}
\let\VANthebibliography\thebibliography
\def\thebibliography{\DeclareRobustCommand{\VAN}[3]{##3}\VANthebibliography}

\usepackage{graphicx}
\usepackage{amsmath}
\usepackage{placeins}
\usepackage{float}
\usepackage{gensymb}
\usepackage{enumitem}
\usepackage{xspace}
\usepackage{ragged2e}

\newcommand{\dvsg}{\texorpdfstring{$\Delta V_{\star-g}$}{Delta V_star-g}\xspace}

\title[Introducing \dvsg]{Introducing \dvsg: a new universal kinematic disturbance parameter}

\author[J. M. Powley et al.]
{Jonah M. Powley,$^{1, 2, 3}$\thanks{E-mail: jmp258@cam.ac.uk (JMP)}
Rebecca J. Smethurst$^{3}$, Chris J. Lintott$^{3}$, Tobias G\'eron$^{4}$\\
$^{1}$Kavli Institute for Cosmology, University of Cambridge, Madingley Road, Cambridge, CB3 0HA, UK\\
$^{2}$Cavendish Laboratory, University of Cambridge, 19 JJ Thomson Avenue, Cambridge, CB3 0HE, UK\\
$^{3}$Oxford Astrophysics, Department of Physics, University of Oxford, Denys Wilkinson Building, Keble Road, Oxford OX1 3RH, UK\\
$^{4}$Dunlap Institute for Astronomy \& Astrophysics, University of Toronto, 50 St George Street, Toronto, ON M5S 3H4, Canada}

\date{Accepted XXX. Received YYY; in original form ZZZ}

\pubyear{\the\year{}}

\begin{document}
\label{firstpage}
\pagerange{\pageref{firstpage}--\pageref{lastpage}}
\maketitle

\begin{abstract}
We introduce a new kinematic disturbance parameter, \dvsg (pronounced `DVSG'), which takes advantage of integral field spectroscopy (IFS) to quantify differences between a galaxy's stellar and gas velocity maps. The motivation behind \dvsg is to capture disturbances in the kinematics of a galaxy that might be missed by alternative methods, while also attempting to minimize bias towards galaxy properties or features of the IFS data. We first detail the reasons for introducing this parameter, and explain how the \dvsg value of a galaxy can be calculated. We then present initial results using \dvsg to quantify the kinematic disturbance of obscured active galactic nuclei (AGN) found in the Mapping Nearby Galaxies at Apache Point Observatory (MaNGA) survey. We find that there is no statistically significant difference between the \dvsg distributions of AGN and a control sample (matched in mass and redshift) of inactive galaxies. This suggests that AGN triggering may not be preferentially caused by any distinct kinematic disturbance process, or combination of processes, beyond those observed in inactive galaxies.
\end{abstract}

\begin{keywords}
galaxies: kinematics and dynamics -- galaxies: active -- galaxies: evolution
\end{keywords}

\section{Introduction}\label{introduction}

Ever since the discovery that supermassive black holes (SMBHs) co-evolve with various properties of their host galaxies \citep[e.g.,][]{magorrian_demography_1998, tremaine_slope_2002, haring_black_2004, kormendy_coevolution_2013}, galaxy mergers have emerged as a plausible mechanism for SMBH growth \citep[e.g.,][]{hopkins_unified_2006}. However, the exact connection between galaxy mergers and SMBH growth, or the triggering of active galactic nuclei (AGN), is still unclear. While galaxy mergers have been shown to trigger AGN \citep[e.g.,][]{alonso_active_2007, ellison_galaxy_2011, gabor_comparison_2016, ellison_definitive_2019}, answering whether mergers are the dominant process behind AGN triggering is less straightforward. Some studies find that AGN are more likely to be found in interacting systems, as well as reside in host galaxies which exhibit greater degrees of morphological disturbance than controls \citep[e.g.,][]{ellison_definitive_2019, pierce_agn_2022}, whereas other studies do not \citep[e.g.,][]{dunlop_quasars_2003, gabor_active_2009, cisternas_bulk_2010, mechtley_most_2016}. Moreover, evidence from observations \citep{cisternas_bulk_2010, simmons_supermassive_2017, smethurst_kiloparsec-scale_2021, garland_galaxy_2024} and simulations \citep{martin_normal_2018, mcalpine_galaxy_2020, smethurst_evidence_2024} suggests that non-merger processes such as secular evolution or disc instabilities may play a significant, even dominant, role in SMBH growth.

Photometric identification of close pairs and interacting galaxies has long been used to study galaxy mergers \citep[e.g.,][]{de_propris_millennium_2007, robotham_galaxy_2014}; however, it cannot provide a complete view of this process. Morphological disturbances, such as tidal features, are transient, and often require deep imaging to be detected \citep{sani_agnstarburst_2014, ellison_definitive_2019}, restricting the maximum timescale after which merger signatures can be observed \citep{lotz_galaxy_2008}. Moreover, simulations suggest that galaxies can form or regrow discs following a merger \citep{font_diversity_2017, pontzen_how_2017, sparre_unorthodox_2017}. This complicates the task of visually identifying whether a galaxy has merged in the past since a galaxy could undergo a merger, form or regrow a disc, and still present as not having experienced a merger.

Measurements of kinematics provide different but complementary information on the history of past galaxy interaction \citep{barrera-ballesteros_tracing_2015, li_impact_2020}. Galaxy mergers tend to transfer stellar trajectories from rotation-supported orbits to dispersion-supported orbits via the redistribution of angular momentum. Additionally, mergers can disturb the gas in a galaxy through gravitational torques, shocks and inflows/outflows \citep[e.g.,][]{barnes_fueling_1991, springel_black_2005, sparre_gas_2022}. These disturbances of the stellar and gas velocity fields in a galaxy can be observed on longer post-coalescence timescales than with photometric methods \citep{mcelroy_observability_2022}.

Kinematic offsets can also be introduced between the stellar and gas velocity fields themselves. While these are not universal across merging/merged galaxies \citep{nevin_accurate_2021}, and are also present in the wider galaxy population \citep{barrera-ballesteros_tracing_2015, ristea_sami_2022}, comparing kinematic disturbance between AGN host galaxies and a control sample of inactive galaxies will enable us to probe the processes that grow SMBHs. If AGN host galaxies tend to be more kinematically disturbed, then perhaps a process like a merger plays a special role in growing these SMBHs. However, if AGN hosts are not more kinematically disturbed, then other galaxy evolution processes may play a similarly important role in SMBH growth.

Investigations of galaxy kinematics have been revolutionized by integral field spectroscopy (IFS). A key advantage of IFS methods over photometric ones is that they can reveal observable kinematic signatures of a galaxy, such as stellar discs, at a large range of inclinations \citep[see Section 3.1 of][]{cappellari_structure_2016}, leading to the possibility of kinematic classification methods that apply to large fractions of the galaxy population \citep{cappellari_sauron_2007, emsellem_sauron_2007}. 

Several parameters have been devised that exploit this capacity of IFS data for robust kinematic classification to probe the kinematic disturbance of galaxies. One such parameter is the difference between the kinematic position angle (PA), or the azimuthal angle from the projected major axis of the plane of the galaxy, of the stellar and gas velocity maps. An approach for computing kinematic PAs is the \textsc{PaFit}\footnote{\url{https://www-astro.physics.ox.ac.uk/~cappellari/software}} package, which minimizes the fit of a symmetrized velocity field to the observed velocity map \citep[see Appendix C of][]{krajnovic_kinemetry_2006}. The kinematic disturbance of a galaxy is then quantified as the absolute difference between the stellar PA and the gas PA, $\Delta \text{PA} = |\text{PA}_{\text{stars}} -\text{PA}_{\text{gas}}|$. $\Delta \text{PA}$ can be computed for a number of different systems, making it a potentially powerful probe of kinematic disturbance \citep[e.g.,][]{krajnovic_atlas3d_2011, ilha_first_2019, ristea_sami_2022}; however, it cannot be computed for all galaxies. There are two requirements for computing a PA of a velocity field: a bi-antisymmetric velocity distribution and a global velocity gradient \citep{krajnovic_kinemetry_2006}. For galaxies that have experienced large kinematic disturbance these assumptions cannot necessarily be made, meaning that $\Delta \text{PA}$ cannot always be accurately computed for these systems. This biases $\Delta \text{PA}$ towards quantifying kinematic disturbance in galaxies with ordered rotation, thereby limiting its universality.

Another kinematic parameter is a proxy for the flux-weighted specific stellar angular momentum, $\lambda_{R_e}$ \citep{emsellem_sauron_2007}. $\lambda_{R_e}$ is a commonly used parameter \citep[e.g.,][]{smethurst_sdss-iv_2018, ristea_sami_2022, zheng_manga_2023}, and has been shown to allow a robust kinematic classification of early-type galaxies (ETGs) as either a dispersion-supported slow rotator or a rotation-supported fast rotator \citep{emsellem_sauron_2007, emsellem_atlas3d_2011, cappellari_structure_2016}. However, despite this success, $\lambda_{R_e}$ faces two issues when it comes to parametrizing kinematic disturbance. Firstly, since $\lambda_{R_e}$ is a flux-weighted value, it will generally be biased towards brighter spaxels in the centre of a galaxy, thereby limiting the contribution of kinematic information around the edges of galaxies, such as tidal features, towards any quantification of kinematic disturbance. Moreover, since $\lambda_{R_e}$ only takes into account a galaxy's stellar velocity, it neglects any kinematic disturbance to the gas in a galaxy.

We therefore propose that other kinematic disturbance parameters should be devised that:
\begin{enumerate}[label=\arabic*., leftmargin=*, listparindent=\parindent]
    \item Draw together the strengths of each of these parameters, probing differences between the stellar and gas kinematics in the case of $\Delta \text{PA}$, and offering a clear kinematic classification of galaxies in the case of $\lambda_{R_e}$.
    \item Avoid the biases of these methods, towards particular galaxy morphologies or less luminous regions of a galaxy for $\Delta \text{PA}$ and $\lambda_{R_e}$, respectively.
\end{enumerate}

We present one such parameter here. This paper proceeds as follows. In Section 2, we outline the data sources of our investigation and formally define our new kinematic disturbance parameter, \dvsg. We present initial results using \dvsg in Section 3, and offer a discussion of these findings in Section 4. Throughout this paper, all uncertainties on quoted median values are the 16th and 84th percentiles of the distribution.

\section{Data and methods}\label{data_methods}

\subsection{MaNGA survey}

To explore the kinematic disturbance of galaxies, we use velocity maps provided by the Sloan Digital Sky Survey-IV (SDSS-IV) Mapping Nearby Galaxies at Apache Point Observatory (MaNGA) survey \citep{bundy_overview_2015} data analysis pipeline \citep[DAP;][]{westfall_data_2019, belfiore_data_2019}. MaNGA is a multi-object survey of $\sim10000$ nearby galaxies ($0.01 < z < 0.15$) which uses 17 integral field units (IFUs), made up of tightly-packed optical fibre bundles, to obtain spatially-resolved spectra across the face of the galaxy. MaNGA samples targets with $M_{\ast} > 10^{9} M_{\odot}$ over an approximately flat distribution in stellar mass, across a wide range of environments, and makes no cuts based on size or inclination. To obtain spectra of these targets, MaNGA uses the Baryon Oscillation Spectroscopic Survey (BOSS) spectrograph \citep{smee_multi-object_2013}, which provides uninterrupted coverage between 3600–10300 Å at a spectral resolution $R \sim 2000$ with an instrumental width $\sim 70 \text{ km s}^{-1}$ \citep[median value $\sim 72 \rm~{km~s}^{-1}$;][]{law_data_2016, chattopadhyay_sdss-iv_2024} across most of this wavelength range.

All spectra are extracted, pre-processed, wavelength and flux calibrated, and astrometrically registered by the MaNGA data reduction pipeline \citep[DRP;][]{law_data_2016} to produce reduced data cubes. These data cubes are then passed to the MaNGA DAP, which provides fits for the continuum, absorption lines and emission lines. From this, two-dimensional maps of the stellar and gas velocity are constructed from the absorption and emission line fits, respectively. We refer the reader to \citet{law_data_2016} for more detailed information about the MaNGA DRP, and \citet{westfall_data_2019, belfiore_data_2019} for more detailed information about the MaNGA DAP.

\subsection{AGN and control sample selection}

To investigate the kinematic disturbance of galaxies and AGN triggering, we study obscured AGN (i.e., with no broad emission lines visible in optical wavelengths) found by MaNGA. We limit our study to obscured AGN since the broad lines of unobscured AGN can dominate the optical spectra of the galaxy. This would lead to inaccurate fits by the MaNGA DAP, which only use a single Gaussian component. Focusing only on obscured AGN will therefore offer us the best data quality to probe kinematic disturbance.

To select a sample of obscured AGN, we make use of the MaNGA AGN catalogue provided by \citet{comerford_excess_2024}, which selects AGN on the basis of mid-infrared colour cuts, hard X-ray sources, 1.4 GHz radio sources, and broad emission lines. Starting from 387 AGN, we apply a conservative quality cut, using the MaNGA data reduction pipeline \citep[DRP;][]{law_data_2016} to only select AGN with the data cube quality flag, \texttt{drp3qual}, equal to 0. From this, we then select AGN that are not flagged either as being X-ray detected or as having broad emission lines so as to eliminate unobscured AGN. After inspecting the SDSS images and central spaxel spectra of each remaining AGN, two additional galaxies (MaNGA PLATEIFUs 8446-1901 and 9181-12702) were removed due to the presence of broad lines that were not flagged previously. Additionally, four AGN (PLATEIFUs 10493-6103, 11020-9101, 10497-12702, 10498-6101) were removed because the presence of other objects in their IFU frames affected the calculation of their \dvsg values (see Section \ref{dvsg_caveats}). This leaves us with a final sample of 209 obscured AGN.

To explore differences between the AGN and other galaxies, we constructed a control sample matched in redshift and stellar mass using data from the MaNGA DRP. First, all MaNGA galaxies were cross-matched within $1^{\prime\prime}$ with the catalogue of bulge and disc decompositions for SDSS galaxies provided by \citet{simard_luc_catalog_2011} as we use this data from this catalogue later on. Additionally, we filtered all remaining galaxies by the same \texttt{drp3qual} quality cut as the AGN, and removed any galaxies with other objects inside the IFU frame significantly affecting the stellar and gas velocity maps. We then used the Hungarian Method \citep{kuhn_hungarian_1955} to find the best matches to the AGN sample in redshift and stellar mass. Testing the robustness of this control sample with an Anderson--Darling two-sample test \citep[AD test;][]{anderson_asymptotic_1952}, we find that the AGN and control sample are unlikely to be drawn from different distributions to high statistical significance in both redshift ($\rm{AD} = -1.21, p = 1.00$) and stellar mass ($\rm{AD} = -1.01, p = 0.98$), suggesting that the control sample is well-matched.

\subsection{Data sources and collection}

From the MaNGA AGN catalogue of \citet{comerford_excess_2024}, we make use of the \texttt{MANGA\textunderscore ID}s of our obscured AGN. We use the NASA-Sloan Atlas to access stellar masses from K-correction fits for elliptical Petrosian fluxes as well as heliocentric redshifts for all AGN and control galaxies. We obtain galaxy bulge fraction, $(B/T)_{r}$, data from the catalogue of bulge and disc decompositions for SDSS galaxies provided by \citet{simard_luc_catalog_2011}. Finally, we use the MaNGA DAP to access the Voronoi binned\citep[][target signal-to-noise ratio = 10]{cappellari_adaptive_2003} stellar and gas velocity maps for all the AGN and control galaxies, as well as their associated masks and inverse variance maps. Additionally, we made extensive use of the Marvin software to access the
MaNGA data \citep{cherinka_marvin_2019}.

\subsection{A new kinematic disturbance parameter: \dvsg} \label{dvsg_introduction}

In this section, we introduce a new parameter, \dvsg (pronounced `DVSG'), which measures the kinematic disturbance in a galaxy. \dvsg is a spatially discretised approach that works by computing the mean absolute difference between the normalized stellar velocity and gas velocity across all bins of a galaxy's velocity map.

We take measures to minimise potential bias towards particular regions of a galaxy or galaxy morphologies. We choose not to flux-weight \dvsg since this would likely bias the parameter towards spaxels closer to the centre of the galaxy, meaning that it could miss information about kinematic disturbance at the edges of the galaxies. We also do not incorporate any information about a galaxy's velocity dispersion, $\sigma$, into \dvsg because the BOSS spectrograph that MaNGA uses has an instrumental width of $\sim 70 \text{ km s}^{-1}$, meaning that it is challenging to robustly constrain dispersions below this scale. Although such dispersions can be inferred with careful modelling of the instrumental line-spread function \citep[e.g.,][]{law_sdss-iv_2021, chattopadhyay_sdss-iv_2024}, doing so in a uniform manner across our sample is non-trivial. Including information about velocity dispersion into \dvsg would therefore bias the parameter towards dispersion-supported galaxies, since the kinematic disturbance of galaxies with $\sigma \lesssim 70 \text{km s}^{-1}$ may not be measured accurately. Given we are introducing a new technique in this paper, we conservatively limit our parametrization of kinematic disturbance to the stellar and gas velocity maps.

We now outline how to calculate the \dvsg value of a galaxy. As stated above, this is performed using stellar and gas velocity maps from the MaNGA survey; however, this calculation could, in principle, be repeated for galaxies with velocity maps from other IFU surveys, or using another gas tracer.

We apply a series of checks to ensure good data quality for the calculation. First, we use the masks from the MaNGA DAP for a galaxy's stellar and gas velocity maps. Only data unmasked in both maps are used. We also add a signal-to-noise ratio (SNR) cut, requiring that any bin used in the \dvsg calculation must have a mean g-band weighted SNR greater than 10. We explored different criteria for this threshold; however, this value offers us the best trade-off between high data quality and availability of data to use in the calculation. An alternative approach is to apply a SNR cut using the gas flux. This can be especially important for ETGs, which can have high-SNR continua but little to no ionized-gas emission. Thus, other authors using \dvsg might consider this cut, especially if their samples contain many ETGs. However, in this work, we continue with the SNR threshold on the continuum. Following this step, we apply a three-sigma clip to both stellar and gas velocity maps to further reduce the sensitivity to outliers or bins with spurious velocity values.

For the calculation itself, we min-max normalize each stellar and gas velocity map so that its values range between -1 and 1, where these extrema correspond to the most negative and most positive velocities relative to the galaxy's systemic velocity, respectively. These maps are normalized individually since we are primarily interested in the offset of the gas velocity from the stellar velocity rather than differences in the magnitudes of the stellar velocity and gas velocity. Next, we compute the absolute difference between each bin in the stellar velocity map and the corresponding bin in the gas velocity map, $|V^{j}_{\star,\text{norm}} - V^{j}_{g,\text{norm}}|$, where $V^{j}_{\star,\text{norm}}$ and $V^{j}_{g,\text{norm}}$ are the normalized stellar and gas velocities in the $j$th bin, respectively. Finally, we take the mean of these residuals to obtain the \dvsg value. Written as an equation, the \dvsg value for a given galaxy is as follows:

\begin{equation}
\Delta V_{\star-g} = \frac{1}{N}\sum_{j=1}^{N} \left| V^{j}_{\star,\mathrm{norm}} - V^{j}_{\mathrm{gas},\mathrm{norm}} \right|
\label{dvsg_equation}
\end{equation}

where $\sum_{j=1}^{N} {|V^{j}_{\star,\text{norm}} - V^{j}_{g,\text{norm}}|}$ denotes the sum over all bins, $j$, of the absolute difference between the normalized stellar and gas velocities. The calculation is implemented in the publicly available Python package \textsc{dvsg} (see Code Availability). For details about how to compute uncertainties on \dvsg values, we refer the reader to Appendix \ref{app:dvsg_uncertainties}.

When interpreting \dvsg, galaxies whose stellar and gas velocity fields are less similar, indicative of kinematic disturbance, tend to exhibit larger \dvsg values. A value of \dvsg = 0 corresponds to identical stellar and gas velocity field distributions. In principle, the maximum possible value of \dvsg is 2, which would occur in the extreme case where all corresponding spatial bins in the two maps take values of 1 and -1, respectively. In practice, however, galaxy kinematics are usually characterised by a mix of positive and negative velocity components that effectively mitigate against large differences. This means that while individual bins or regions of a galaxy can approach a residual value of 2, galaxies have \dvsg values much lower than this. Both our analysis and testing show that \dvsg values rarely exceed 1 for galaxies in this study.

\begin{figure*}
    \includegraphics[width=\textwidth]{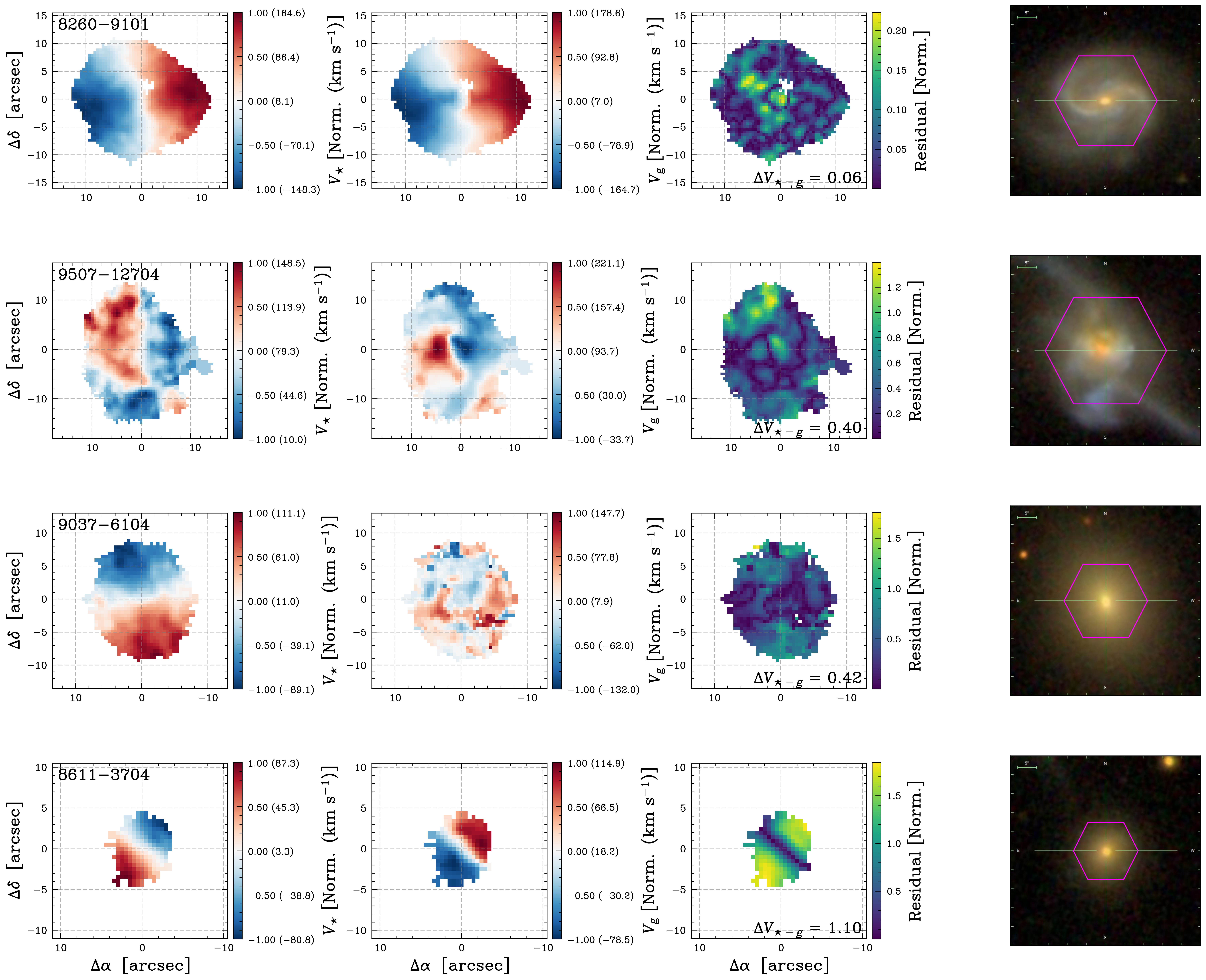}
    \caption{Illustration of \dvsg. Left column: Stellar velocity maps for four MaNGA galaxies (PLATEIFUs shown in the top-left corner for reference), showing the normalized and unnormalized stellar velocity values in the colour bar ticks of each velocity map, with the latter enclosed in brackets. Second column: Gas velocity maps for the same galaxies, also showing normalized and unnormalized velocity values indicated by the colour bar ticks. Third column: Maps of the absolute difference between the normalized stellar and gas velocity for each galaxy, with the \dvsg value of the galaxy shown in the bottom-right corner. Right column: SDSS images of the galaxies, overlaid with the MaNGA IFU field of view. All maps shown are processed following the procedure in Sec. \ref{dvsg_introduction}.}
    \label{fig:dvsg_example_calculation}
\end{figure*}

This is illustrated in Fig. \ref{fig:dvsg_example_calculation}, which shows how the \dvsg values of four MaNGA galaxies have been calculated. The galaxy in the top panel has stellar and gas velocity maps that are both rotation-supported with little difference between them, leading to a low \dvsg value. The galaxy in the upper-middle panel exhibits reasonably symmetric stellar and gas velocity maps; however, discrepancies between the two, highlighted in the residual map, indicate kinematic disturbance. This disturbance is likely driven by the ongoing galaxy merger visible in the SDSS image. Consequently, this galaxy has a larger \dvsg value than the galaxy shown in the upper panel. The galaxy in the lower-middle panel has a rotation-supported stellar velocity field, but a more disturbed gas velocity field, which also leads to a higher \dvsg value. Finally, the galaxy in the bottom panel has stellar and gas velocity maps that are both rotation-supported but are counter-rotating ($\Delta \text{PA} \sim 180\degree$), resulting in a near-maximally high \dvsg value.

\begin{figure}
    \includegraphics[width=\columnwidth]{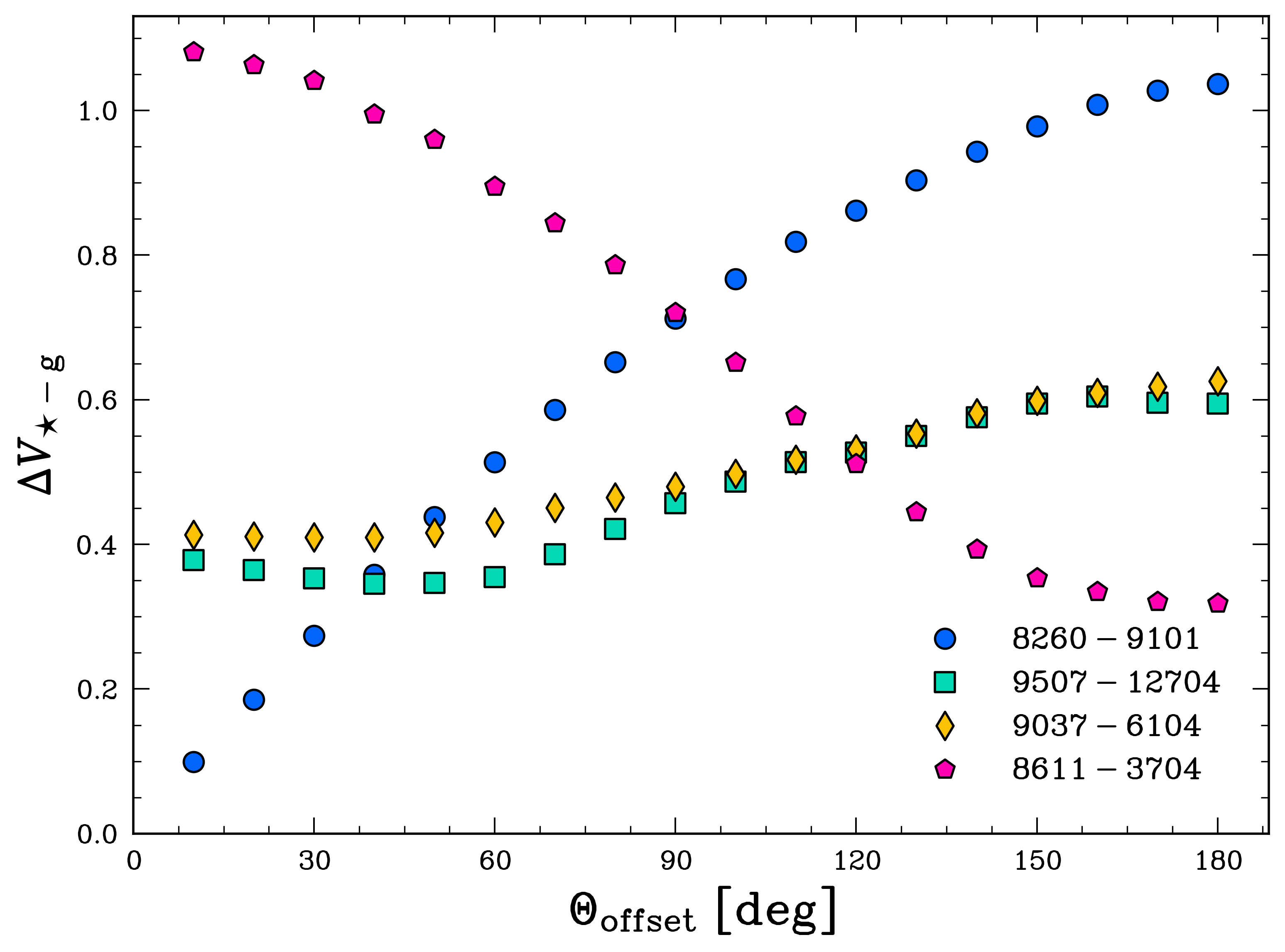}
    \caption{Scatter plot of \dvsg values for the four galaxies shown in Fig. \ref{fig:dvsg_example_calculation} as a function of the artificial rotation angle of the gas velocity map. MaNGA PLATEIFUs in the legend indicate the galaxy to which each set of markers correspond.}
    \label{fig:dvsg_galaxy_rotation_test}
\end{figure}

To demonstrate how a galaxy's \dvsg value would change for different kinematic offsets between the stellar and gas components, we calculate \dvsg values for the four example galaxies in Fig. \ref{fig:dvsg_example_calculation} after artificially rotating their gas velocity maps by angles between 0 and 180 degrees. Because the MaNGA data are Voronoi-binned, this is achieved by rotating the spaxel-level gas velocity map while preserving the spatial coordinates of each bin. Following each artificial rotation, we calculate \dvsg using the mean residual between the stellar velocity field and the rotated gas velocity field within each bin. As a validation step, we confirm that the artificial rotation preserves the gas velocity distribution relative to the original map. This means that we have created a set of realistic velocity maps at a range of offset angles allowing us to test how \dvsg traces kinematic disturbance. The results of this test are shown in Fig. \ref{fig:dvsg_galaxy_rotation_test}.

The \dvsg values of galaxies with rotation-supported kinematics (top and bottom panels in Fig. \ref{fig:dvsg_example_calculation}) can change significantly as a function of offset angle, going from around 1 when the stellar and rotated gas maps are anti-aligned to between 0.1 and 0.3 when they are aligned. Moreover, when the bottom (initially counter-rotating) galaxy in Fig. \ref{fig:dvsg_example_calculation} is artificially rotated close to 180 degrees, the galaxy's \dvsg value approaches 0.3. On the other hand, galaxies which show a greater degree of disturbed kinematics (middle panels of Fig. \ref{fig:dvsg_example_calculation}) do not see the same variability, with their \dvsg values, ranging between 0.3 and 0.6. Although these tests only show how \dvsg changes as a function of the artificial offset angle for four galaxies, it suggests that galaxies with aligned, rotation-supported stellar and gas kinematics can be distinguished from all other combinations of rotation-supported or dispersion-supported stellar and gas velocity profiles on the basis of their \dvsg value. To ensure that our conclusions are not specific to the observed galaxies, we repeat the same analysis using toy models of both rotation-supported and dispersion-supported velocity fields and find very similar results.

\section{Results}\label{results}

\begin{figure}
    \includegraphics[width=\linewidth]{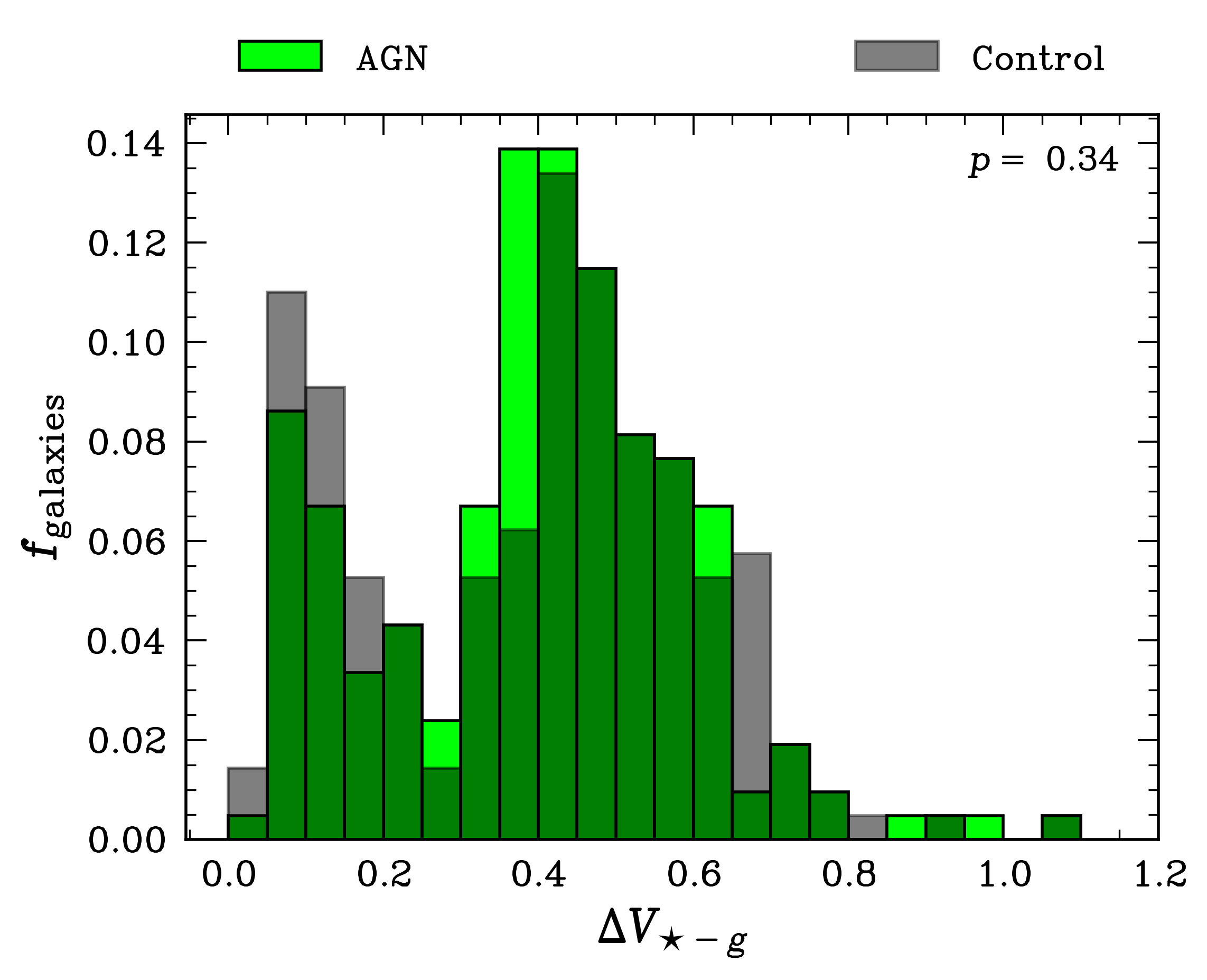}
    \caption{Histogram plots of the \dvsg distribution for the AGN (green) and control galaxies (grey). The AD test p-value is shown in the top right.}
    \label{fig:dvsg_distributions_agn_control}
\end{figure}

Fig. \ref{fig:dvsg_distributions_agn_control} shows the distributions of \dvsg for the AGN and inactive control galaxies. An AD test ($\rm{AD} = 0.01, p=0.35$) indicates that we cannot reject the null hypothesis that the AGN and control samples have statistically indistinguishable kinematic disturbance. Since we find the \dvsg distributions of AGN and control galaxies to be statistically indistinguishable, it suggests that AGN-host galaxies are unlikely to be affected by distinct kinematic disturbance processes, or combinations of processes, that are not also present in inactive control galaxies.

\begin{figure}
    \centering
    \includegraphics[width=\columnwidth]{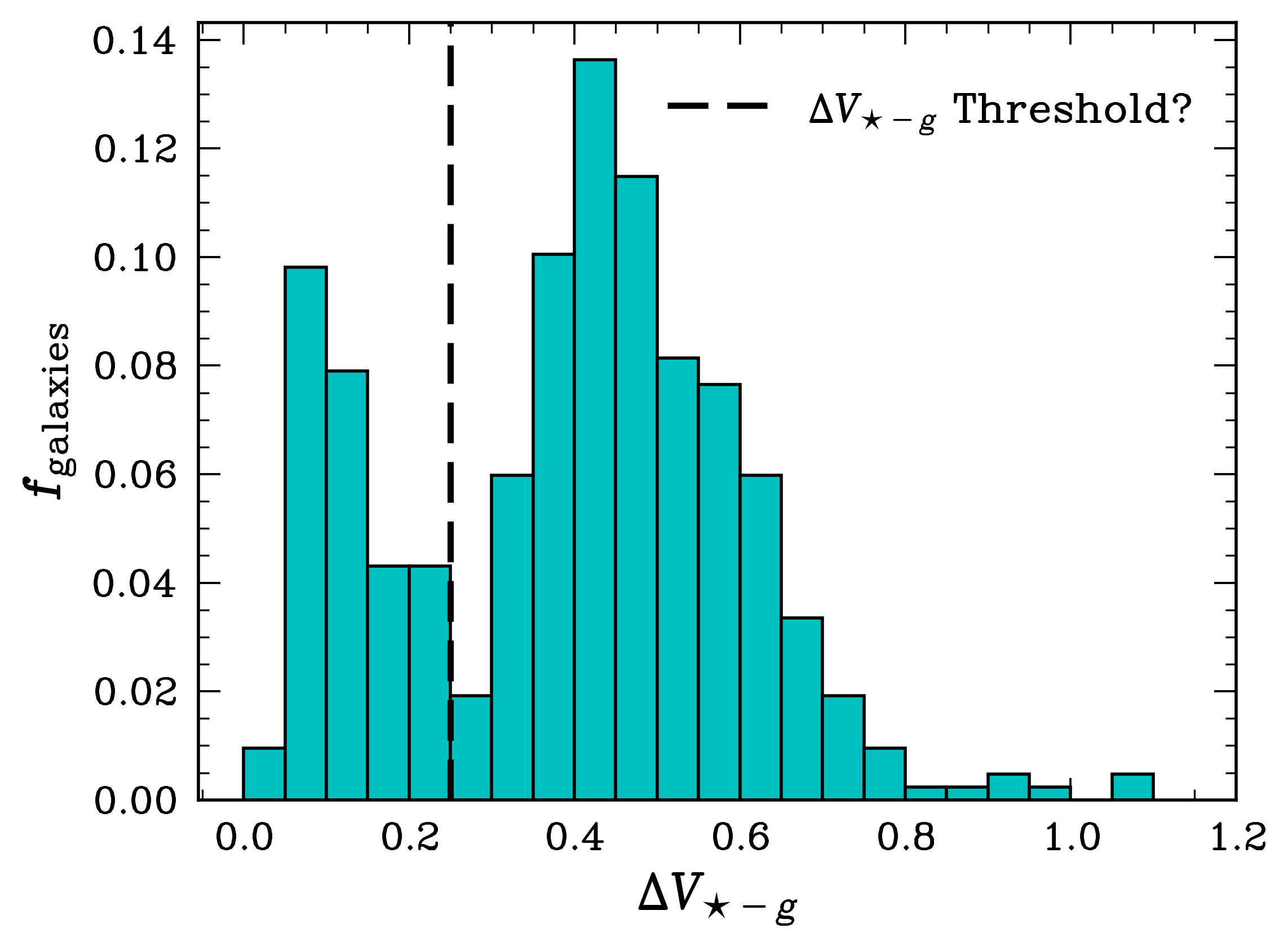}
    \caption{Histogram plot of the \dvsg distribution for all galaxies (AGN and control) in our sample. The black dashed line shows a possible threshold at \dvsg = 0.25, below which we find galaxies with aligned, rotation-supported stellar and gas velocity maps, and above which significantly fewer galaxies exhibit this kinematic morphology.}
    \label{fig:dvsg_distribution_all_galaxies}
\end{figure}

We show the \dvsg distribution for all galaxies in our sample in Fig. \ref{fig:dvsg_distribution_all_galaxies}. We note that this distribution is bi-modal, with peaks at $\sim$ 0.1 and 0.5, and a minimum between them at $\sim$ 0.25. From visual inspection of our sample, we find that all galaxies below a \dvsg value of $\sim0.25$ show clear rotation-supported kinematics and aligned stellar and gas velocity maps. Moreover, above this \dvsg value, we find significantly fewer galaxies whose kinematics match this description. Since these \dvsg values correspond strongly with the minimum of the \dvsg distribution for all galaxies, as well as the \dvsg values of the artificial rotation testing, it suggests that this region of the parameter space is tracing the population of galaxies with aligned, rotation-supported stellar and gas kinematics.

In the interest of investigating this population further, we adopt a threshold in \dvsg of 0.25. We refer to galaxies below this threshold as `kinematically undisturbed' (\dvsg < 0.25), and refer to other galaxies in our sample, which display a set of different kinematic profiles, as `kinematically disturbed' (\dvsg > 0.25). We emphasise here that, although this nomenclature is empirically motivated, it is adopted in part for clarity of presentation and may not align exactly with usage in other subfields. In particular, `kinematically undisturbed' indicates minimal observed kinematic disturbance, but does not rule out some degree of past disturbance. Conversely, `kinematically disturbed' reflects observable irregularities in the stellar and gas kinematics, but does not imply the absence of ordered rotation. Applying this threshold to our sample, we identify 114 galaxies as kinematically undisturbed and 304 galaxies as kinematically disturbed. This threshold is not intended to be an exact constraint. As indicated above, some galaxies with \dvsg > 0.25 show aligned, rotation-supported stellar and gas kinematics. Additionally, by varying the value of the SNR cut or sigma clip before the \dvsg calculation, we find that the threshold value can shift by $\sim0.05$. However, in all these cases, we still find that a similar bi-modality in the \dvsg distribution emerges.

\begin{figure*}
    \includegraphics[width=\textwidth]{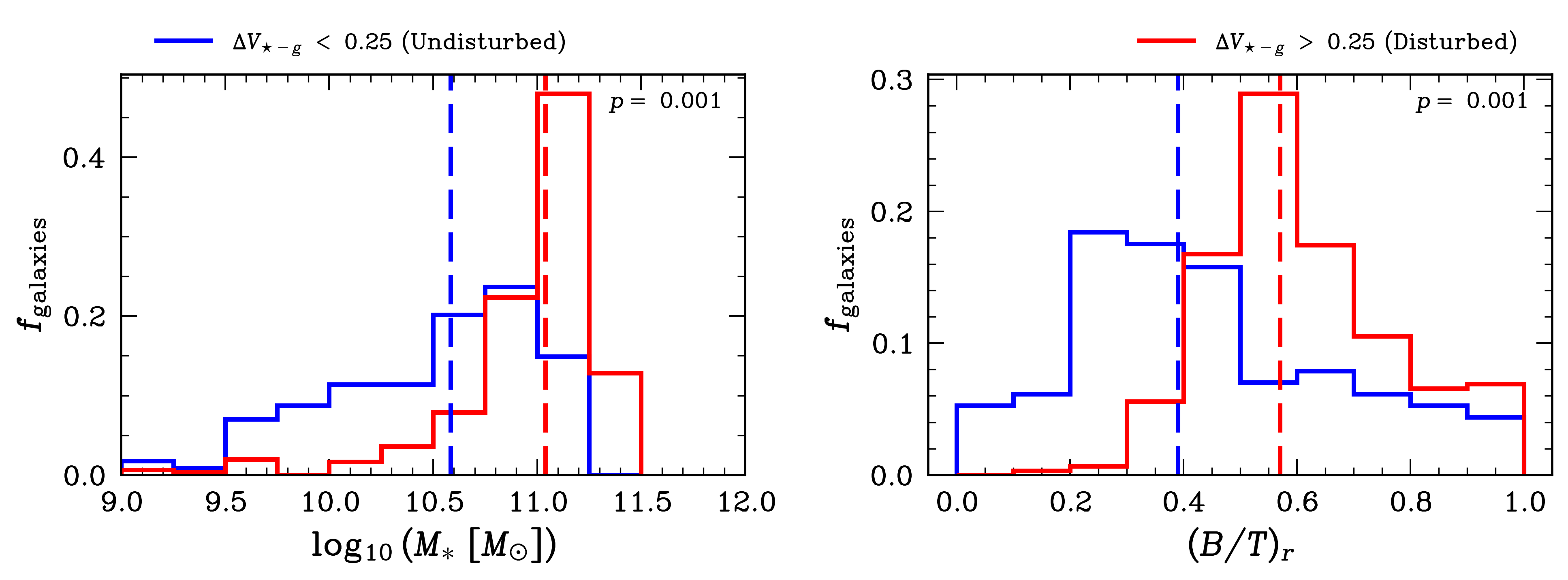}
    \caption{Histograms of $\log_{10}(M_{\ast}/M_{\odot})$ for galaxies classified as kinematically undisturbed (blue) and kinematically disturbed (red). Right: Corresponding histograms of bulge fraction $(B/T)_{r}$ for the same samples, using the same colour scheme. In each panel, dashed lines indicate the median value of each population, and the Anderson–Darling test p-value is shown in the top-right corner.}
    \label{fig:MStellar_And_BTr_Histograms_For_Low_Vs_High_DVSG}
\end{figure*}

The left panel of Fig. \ref{fig:MStellar_And_BTr_Histograms_For_Low_Vs_High_DVSG} shows the distributions of stellar mass for all AGN and control galaxies, separated between those identified as kinematically disturbed or undisturbed. The median stellar mass values for each sample are $10^{10.58_{-0.62}^{+0.40}} M_{\odot}$ and $10^{11.04_{-0.27}^{+0.20}} M_{\odot}$ for the undisturbed and disturbed galaxies, respectively. The p-value of an AD test suggests that we can reject the null hypothesis that kinematically disturbed and undisturbed galaxies are drawn from the same distribution ($\rm{AD} = 63.22, p = 0.001$), meaning that they are statistically distinguishable by their stellar mass. This could suggest that \dvsg is tracing processes that grow a galaxy's stellar mass.

Similarly, the right panel of Fig. \ref{fig:MStellar_And_BTr_Histograms_For_Low_Vs_High_DVSG} shows the distribution of galaxy bulge fraction for all AGN and control galaxies, with the same separation into kinematically disturbed and kinematically undisturbed as the left panel. The median galaxy bulge fraction values for each sample are {$0.39_{-0.16}^{+0.32}$} and $0.57_{-0.11}^{+0.21}$ for the undisturbed and disturbed galaxies, respectively. An AD test of the galaxy bulge fraction distributions suggests that the two populations are statistically distinguishable ($\rm{AD} = 45.86, p=0.001$). This could suggest that \dvsg is tracing processes that grow a galaxy's bulge.

\section{Discussion}

\subsection{Processes that \dvsg could be tracing in kinematically disturbed and undisturbed galaxies}\label{dvsg_tracing_processes}

We argued in Section \ref{results} that \dvsg may trace processes that grow galaxy stellar masses and bulges. We consider one such process to be galaxy mergers. Since mergers are associated with greater kinematic disturbance in a galaxy and are known to increase a galaxy's stellar mass and bulge fraction \citep[e.g.,][]{toomre_galactic_1972, hopkins_mergers_2010, tonini_growth_2017}, we argue that they are the most suitable candidate to explain the observed relationship between a galaxy's kinematic disturbance and its stellar mass. 

This is supported by recent analysis of the Horizon-AGN simulation by \citet{smethurst_evidence_2024}, who found that the most massive black holes, which reside in the most massive galaxies, tend to be grown by mergers. We note that this is consistent with saying that mergers do not preferentially trigger AGN, since our results comparing kinematically disturbed and undisturbed galaxies consider both AGN and control galaxies. Our findings simply suggest that, depending on whether a galaxy is kinematically disturbed or undisturbed, we might expect different processes to be responsible for increasing that galaxy’s stellar mass and galaxy bulge fraction.

An interesting result for kinematically undisturbed galaxies is that, while they are statistically distinguishable from disturbed galaxies with respect to bulge fraction, some undisturbed galaxies have similar bulge fraction values as the disturbed galaxies. Although we cannot rule out the possibility that these may be pseudo-bulges rather than the geometric bulges expected in dispersion-supported and kinematically disturbed galaxies \citep[e.g.,][]{kormendy_kinematics_1993, devergne_bulge_2020}, it is worth considering what processes might be occurring in galaxies identified by \dvsg as kinematically undisturbed to grow such (pseudo-) bulges. A strong candidate process here is disc instabilities. Analysis of the Millennium simulation found that when $M_{\ast} < 10^{11} M_{\odot}$, the vast majority of the stellar mass of the bulge component of the galaxy is grown via instabilities \citep{bower_breaking_2006, parry_galaxy_2009}. Since kinematically undisturbed galaxies have a median stellar mass value ($10^{10.61^{+0.37}_{-0.52}} M_{\odot}$) within the range that the Millennium simulation indicates stellar mass growth occurs mostly via instabilities, while disturbed galaxies have a median stellar mass value just above the $10^{11} M_{\odot}$ threshold ($10^{11.02^{+0.20}_{-0.42}} M_{\odot}$), disc instabilities may represent a plausible mechanism for bulge growth process in kinematically undisturbed galaxies. Though disc instabilities might still affect the kinematics of a galaxy, since this process tends to be less violent than a merger, galaxies grown via instabilities may still be relatively kinematically undisturbed. 

A route for future investigation here is to calculate \dvsg values for mock observations of zoom-in galaxy simulations to allow a more controlled test of kinematic processes that \dvsg might be tracing.

\subsection{The role of kinematic disturbance processes in the triggering of AGN} \label{agn_triggering}

The results presented in Section \ref{results} suggest that AGN may not be preferentially triggered by any distinct kinematic disturbance processes. The reasoning for this is that if there was a process that dominated the triggering of AGN, then we would expect the \dvsg distribution of the AGN sample to cluster around the region of \dvsg values that might be associated with that process compared with control galaxies. For example, if AGN were preferentially triggered by galaxy mergers, then we would expect the AGN sample to generally have higher \dvsg values than control galaxies to reflect the greater kinematic disturbance caused by mergers. Equally, if AGN were preferentially triggered by secular evolution, then we would expect to see the AGN sample tend towards lower \dvsg values than control galaxies to reflect the lesser degree of kinematic disturbance that results from secular evolution. Because we do not see such shifts in the \dvsg distributions of the AGN relative to the control sample – finding the two populations to be statistically indistinguishable – it suggests that, at least approximately, the same set of kinematic disturbance processes that are happening in AGN host galaxies are also at play in control galaxies. Put another way, this result might mean that there is no galaxy evolution process that dominates the growth history of SMBHs.

\citet{ilha_first_2019} investigated the kinematic disturbance of 62 MaNGA AGN, using the catalogue of \citet{rembold_first_2017}, by comparing the $\Delta\text{PA}$ values of their AGN to 109 inactive control galaxies, and found no statistically significant difference between the $\Delta \text{PA}$ distributions of the AGN and control samples. When we calculate \dvsg values for all the AGN and control galaxies used by \citet{ilha_first_2019}, we also determine them to have statistically indistinguishable \dvsg distributions ($\rm{AD} = -0.42, p = 0.60$). Comparing the $\Delta \rm{PA}$ and \dvsg\ values for this sample, we find that the median \dvsg value for all galaxies with $\Delta \rm{PA}$ less than 30 degrees is $0.13^{+0.29}_{-0.07}$, meaning that most of these galaxies would also be classified as kinematically undisturbed using \dvsg = 0.25 (71.25 per cent or 114 / 160). That the results of \citet{ilha_first_2019} agree with our findings about the statistically indistinguishable kinematic disturbance of AGN and control galaxies, and galaxies in their sample with low $\Delta \rm{PA}$ values also tend to have low \dvsg values, corroborates our analysis and demonstrates the success of \dvsg at probing galaxy kinematics.

In another study, \citet{comerford_excess_2024} apply SDSS merger classifications from a model trained on mock images of merging and non-merging simulated galaxies \citep[see][]{nevin_declining_2023} to the MaNGA AGN catalogue used in this work \citet{comerford_excess_2024}. They find that galaxies in MaNGA identified as mergers have an AGN fraction over twice that of galaxies identified as non-mergers ($4.2 \pm 0.3$ per cent versus $1.8 \pm 0.2$ per cent). We note that despite the different approach taken in this paper (comparing the AGN fraction of galaxy mergers and non-mergers, as opposed to comparing the kinematic disturbance of AGN and inactive control galaxies as in this work), the results of \citep{comerford_excess_2024} are not incompatible with our own. Inspecting the AGN in Fig. \ref{fig:dvsg_distributions_agn_control}, it is clear that a larger proportion of this sample are classified as kinematically disturbed rather than kinematically undisturbed. Combined with the inference made in Section \ref{dvsg_tracing_processes} that the disturbed population is more likely to contain merged galaxies than the undisturbed population, this kinematic disturbance excess in AGN can be seen as concordant with the AGN excess in mergers reported by \citet{comerford_excess_2024}. Understanding this result in light of our findings in Section \ref{results} suggests a picture where mergers may be involved in the triggering of AGN, but are not the sole dominant process, perhaps because multiple galaxy evolution mechanisms are jointly required to fuel AGN.

One concern with our results could be the discrepancy between the timescales of the AGN duty cycle and kinematic disturbance. While the activity from an AGN can vary rapidly \citep[$\sim\!1\!-\!100\,{\rm Myr}$, e.g., ][]{hickox_black_2014, harrison_impact_2017}, the timescales of observing kinematic disturbance from galaxy mergers can range between $\sim\!0.25\!-\!1\,\rm{Gyr}$ depending on the merger type and kinematic tracer used \citep{hung_merger_2016}. From this perspective, the results presented in Fig. \ref{fig:dvsg_distributions_agn_control} could be interpreted as reflecting a temporal offset: if a merger event triggers an AGN that subsequently fades by the epoch of observation, the galaxy would retain the kinematic signature of the merger but would no longer be classified as an AGN host. This could dilute any observable connection between merger-induced kinematic disturbance and AGN activity. While it is challenging to quantify this effect of the AGN duty cycle on our sample, we argue that it is unlikely to play a significant role.

Recent evidence suggests that secular processes, which are thought to occur via planar galaxy-scale gas flow \citep{nayakshin_observed_2012}, are sufficient to fuel black hole growth in the local Universe. \citet{smethurst_secularly_2019} studied the inflow and outflow rates of a sample of disc-dominated AGN, which are assumed to have merger-free growth histories since at least $z\sim2$ \citep[][]{martin_normal_2018}. Compared with a merger-dominated sample of AGN, the inflow rates for the disc-dominated sample were $\sim5$ times higher, whereas the outflow rates were $\sim5$ times lower. Since our kinematically undisturbed sample shows regularly rotating, aligned stellar and gas kinematics as well as low B/T ratios, we can assume the AGN have grown via secular processes. We argue it is more likely that these galaxies will continue to grow via secular feeding, and therefore that they will not undergo rapid changes in their AGN variability within the timescales that \dvsg may be sensitive to disturbance.

The same argument cannot necessarily be made for our kinematically disturbed sample. Black hole growth in these galaxies is more likely to be driven by less radiatively-efficient quasi-spherical inflows, which would result in stronger outflows than secularly-grown black holes \citep{nayakshin_observed_2012}. However, we note that the \dvsg distributions of the AGN and control samples are statistically indistinguishable. This could suggest that the processes that are powering kinematically undisturbed galaxies may also be powering kinematically disturbed galaxies, but further studies would be needed to investigate this fully.

As mentioned in Section \ref{data_methods}, the restriction of the MaNGA DAP fitting procedure limits our study to obscured AGN. Future work could therefore extend the calculation of \dvsg to unobscured AGN by refitting the data cubes to include a broad line component to account for unobscured AGN.

\subsection{Caveats and robustness of \dvsg}\label{dvsg_caveats}

In this section, we describe potential caveats of using \dvsg to quantify kinematic disturbance, and evaluate the robustness of the parameter to data preprocessing choices, steady-state dynamical effects, and observational systematics.

As outlined in Sec. \ref{dvsg_introduction}, there are several data preprocessing steps that must be applied to the stellar and gas velocity maps in order to calculate \dvsg, namely masking, sigma-clipping and normalization. We test the effect of using different normalization schemes (e.g., z-score or maximum absolute deviation) in the \dvsg calculation, and find no significant change to our results. Another concern here is the impact of outliers in the velocity map data on the \dvsg value. Image artefacts, unmasked objects in the IFU or inaccurate fits to the galaxy's data cube can all lead to some bins having spuriously high stellar and gas velocity values. Since we normalize the stellar and gas velocity maps between -1 and 1, significantly larger values can suppress values in other bins, resulting in the sum across a galaxy being unrepresentative of the kinematic disturbance. As mentioned above, we mitigate this by employing a SNR cut of 10 (the same target SNR as the MaNGA DAP) and excluding all bins with velocities more than three standard deviations from the mean. These measures generally work well to reduce the influence of velocity map outliers in the majority of our sample; however, some galaxies may need to be dealt with on an individual basis. This can be done, for example, by applying more stringent SNR cuts and sigma clips, or excluding the galaxy from a sample if necessary. We therefore conclude that while \dvsg is moderately sensitive to extreme velocity outliers, standard quality cuts are sufficient for the majority of galaxies. Future work could employ more sophisticated techniques to mask other objects that appear in the IFU frame or alternative outlier removal strategies.

We next consider whether \dvsg may be driven by steady-state dynamical effects rather than genuine kinematic disturbance. One such effect is asymmetric drift (AD), which is caused by the lower circular velocity of the stars in equilibrium disc galaxies compared to the gas, leading to an offset between the stellar and gas velocity maps \citep[e.g.,][]{bershady_asymmetric_2024}. To quantify the effect of AD on \dvsg, we calculate \dvsg values for 7 MaNGA galaxies\footnote{None of which are present in our sample.}, selected by \citet{shetty_stellar_2020} as having extremely regular stellar and gas velocity maps and clear signs of AD from the MaNGA pipeline products. The resulting \dvsg\ values are low, ranging from $\sim0.05$ to $\sim0.08$, with a mean of $\sim0.06$. Because these galaxies have highly regular stellar and gas velocity maps and lack other sources of kinematic disturbance, we argue that AD is the main process that \dvsg is sensitive to in these galaxies. The low \dvsg values therefore suggest that, although we do not apply an explicit AD correction, our results would only be minimally affected by AD.

Finally, we evaluate the influence of observational systematics. The median ratio of the uncertainty on \dvsg to its value is $0.04_{-0.03}^{+0.04}$, while the standard errors of the \dvsg distributions for the kinematically undisturbed and kinematically disturbed samples are 0.005 and 0.008, respectively. This indicates that measurement error or sample variance do not impact our results. We also report no strong correlation of \dvsg with either the number of IFU fibres or the number of bins per galaxy. Together, these checks suggest that \dvsg is robust to these potential observational systematics.

\section{Summary}

In this paper, we set out a new parameter to measure the kinematic disturbance of a galaxy, \dvsg. This parameter is able to trace differences between stellar and gas velocity fields that are not captured by other parameters, while minimizing biases towards any particular kinematic morphologies or region of a galaxy. Obtaining measurements of \dvsg for 209 obscured AGN from the SDSS-IV MaNGA survey, as well as a mass and redshift-matched control sample, we investigated the connection between kinematic disturbance and the triggering of AGN. Key results from this paper are as follows:

\begin{enumerate}[label=(\roman*), leftmargin=0pt, itemindent=2em, listparindent=1em]

\item AGN and control galaxies, as measured by \dvsg, have a statistically indistinguishable distribution in their kinematic disturbance. This suggests that AGN may not be preferentially triggered by any distinct kinematic disturbance processes, a result supported by previous investigations of kinematic disturbance in galaxies.\\

\item Different regions of the \dvsg parameter space contain galaxies with distinct kinematic morphologies. Galaxies with aligned, rotation-supported stellar and gas velocity fields tend to have lower \dvsg values ($\Delta V_{\star-g} < 0.25$), whereas galaxies of other kinematic types, including dispersion-supported or counter-rotating galaxies, have higher \dvsg values ($\Delta V_{\star-g} > 0.25$). We refer to these two galaxy populations with low and high \dvsg values as kinematically disturbed and kinematically undisturbed, respectively.\\

\item The distributions of stellar mass and galaxy bulge fraction for kinematically disturbed and undisturbed galaxies are statistically distinguishable ($>3 \sigma$ significance). This suggests that \dvsg can trace kinematic disturbance processes that grow stellar masses and galaxy bulge structures for these two populations. Drawing on results from galaxy simulations, we argue that a plausible process for triggering AGN in kinematically disturbed galaxies is galaxy mergers, whereas in kinematically undisturbed galaxies a possible candidate is disc instabilities.
\end{enumerate}

As mentioned above, there are opportunities for further progress using both observations and simulations. Refitting the MaNGA data cubes could extend calculations of \dvsg to unobscured AGN. Moreover, calculating \dvsg values for simulated IFU kinematics could allow for a more accurate understanding of the processes that \dvsg is tracing. We end by noting that \dvsg has a wide range of potential applications within the field of galaxy evolution, and encourage others to use this parameter in future studies.

\section*{Acknowledgements}

We thank the referee, Kyle Westfall, for the helpful comments that greatly improved the paper. JMP would like to thank A. Lola Danhaive for insightful conversations about kinematics. JMP is grateful to the Cambridge Trust and Sidney Sussex College, Cambridge for a Vice Chancellor's \& Sidney Sussex Basil Howard PhD scholarship. RJS gratefully acknowledges funding from the Royal Astronomical Society. TG is a Canadian Rubin Fellow at the Dunlap Institute. The Dunlap Institute is funded through an endowment established by the David Dunlap family and the University of Toronto.

Funding for the Sloan Digital Sky Survey IV has been provided by the Alfred P. Sloan Foundation, the U.S. Department of Energy Office of Science, and the Participating Institutions. SDSS acknowledges support and resources from the Center for High-Performance Computing at the University of Utah. The SDSS web site is \url{www.sdss4.org}. SDSS is managed by the Astrophysical Research Consortium for the Participating Institutions of the SDSS Collaboration including the Brazilian Participation Group, the Carnegie Institution for Science, Carnegie Mellon University, Center for Astrophysics | Harvard \& Smithsonian (CfA), the Chilean Participation Group, the French Participation Group, Instituto de Astrof\'isica de Canarias, The Johns Hopkins University, Kavli Institute for the Physics and Mathematics of the Universe (IPMU) / University of Tokyo, the Korean Participation Group, Lawrence Berkeley National Laboratory, Leibniz Institut f\"ur Astrophysik Potsdam (AIP), Max-Planck-Institut f\"ur Astronomie (MPIA Heidelberg), Max-Planck-Institut f\"ur Astrophysik (MPA Garching), Max-Planck-Institut f\"ur Extraterrestrische Physik (MPE), National Astronomical Observatories of China, New Mexico State University, New York University, University of Notre Dame, Observat\'orio Nacional / MCTI, The Ohio State University, Pennsylvania State University, Shanghai Astronomical Observatory, United Kingdom Participation Group, Universidad Nacional Aut\'onoma de M\'exico, University of Arizona, University of Colorado Boulder, University of Oxford, University of Portsmouth, University of Utah, University of Virginia, University of Washington, University of Wisconsin, Vanderbilt University, and Yale University.

This work made use of Astropy:\footnote{http://www.astropy.org} a community-developed core Python package and an ecosystem of tools and resources for astronomy \citep{the_astropy_collaboration_astropy_2013, the_astropy_collaboration_astropy_2018}, and NASA's Astrophysics Data System, funded by NASA under Cooperative Agreement 80NSSC21M00561.

\section*{Data Availability}

All data used in this paper are available upon reasonable request to the first author.

\section*{Code availability}

The package, \textsc{dvsg}, created for this work to calculate \dvsg, is publicly available at
\url{https://github.com/jonahpowley/dvsg}.

\appendix

\section{Uncertainties on \dvsg}\label{app:dvsg_uncertainties}

In this Appendix, we briefly detail the calculation of uncertainties for \dvsg values. First, we make explicit the min-max normalisation of the stellar and gas velocity maps following the sigma clip:

\begin{equation}\label{dvsg_normalisation}
V^{j}_{\rm{norm}}
= 2 \frac{V^{j} - \max({V^j})}{\max({V^j}) - \min({V^j})} - 1
\end{equation}

where $V^{j}$ refers to the velocity in the $j$th bin of either the stellar or gas velocity map. As mentioned in Sec. \ref{dvsg_introduction}, each velocity map is individually normalised (i.e., with respect to its own minimum and maximum values). Under the assumption of independent bins and fixed min–max bounds, the uncertainty on \dvsg values can be calculated in quadrature using standard error propagation:

\begin{equation}\label{dvsg_error}
\sigma(\Delta V_{\star-g})
= \frac{1}{N}\sqrt{\sum_{j=1}^{N}
\left[
\left(\frac{2}{R_\star}\,\sigma(V^j_\star)\right)^2
+
\left(\frac{2}{R_{\rm{g}}}\,\sigma(V^j_{\rm{g}})\right)^2
\right]}
\end{equation}

Here $\sigma(V^j_\star)$ and $\sigma(V^j_{\rm g})$ represent the uncertainties of the unnormalized stellar and gas velocity map bins, respectively, while $R_\star$ and $R_{\rm g}$ are the (sigma-clipped) velocity ranges used for the min–max normalization in equation \ref{dvsg_normalisation}. The factors of $2/R_{\star}$ and $2/R_{\rm{g}}$ arise from rescaling by the normalized interval length. We note that MaNGA bins or spaxels are known to be correlated; however, incorporating such covariance into the \dvsg error calculation is non-trivial and we do not expect this to dramatically change our results.

\bibliographystyle{mnras}

\bibliography{DVSG_ADS}

\bsp
\label{lastpage}
\end{document}